\documentclass[aps, prl, 
amsmath, showpacs, preprintnumbers, twocolumn, amssymb, amsmath, superscriptaddress]{revtex4-1}
\usepackage{graphicx}
\usepackage{mathptmx, textcomp}
\usepackage[latin1]{inputenc}
\usepackage{tabularx}
\usepackage{units}
\usepackage{color}
\usepackage{bm}

\begin{document}
\title{Quantum squeezing of a levitated nanomechanical oscillator}
\author{M.\,Kamba}
\affiliation{Department of Physics, The University of Tokyo;  7-3-1 Hongo, Bunkyo-ku, 113-0033 Tokyo}
\author{N.\,Hara}
\affiliation{Department of Physics, The University of Tokyo;  7-3-1 Hongo, Bunkyo-ku, 113-0033 Tokyo}
\author{K.\,Aikawa}
\affiliation{Department of Physics, The University of Tokyo;  7-3-1 Hongo, Bunkyo-ku, 113-0033 Tokyo}

\date{\today}


\begin{abstract}
Manipulating the motions of macroscopic objects near their quantum mechanical uncertainties has been desired in diverse fields, including fundamental physics, sensing, and transducers. Despite significant progresses in ground-state cooling of a levitated solid particle, realizing non-classical states of its motion has been elusive. Here, we demonstrate quantum squeezing of the motion of a single nanoparticle by rapidly varying its oscillation frequency. We reveal significant narrowing of the velocity variance to $-4.9(1)$~dB of that of the ground state via free-expansion measurements. To quantitatively confirm our finding, we develop a method to accurately measure the displacement of the nanoparticle by referencing an optical standing wave. Our work shows that a levitated nanoparticle offers an ideal platform for studying non-classical states of its motion and paves the way for its applications in quantum sensing, as well as for exploring quantum mechanics at a macroscopic scale.
\end{abstract}

\maketitle

%Introduction

\begin{figure} % Do NOT use \begin{figure*}
	\centering
	\includegraphics[width=0.9\columnwidth]{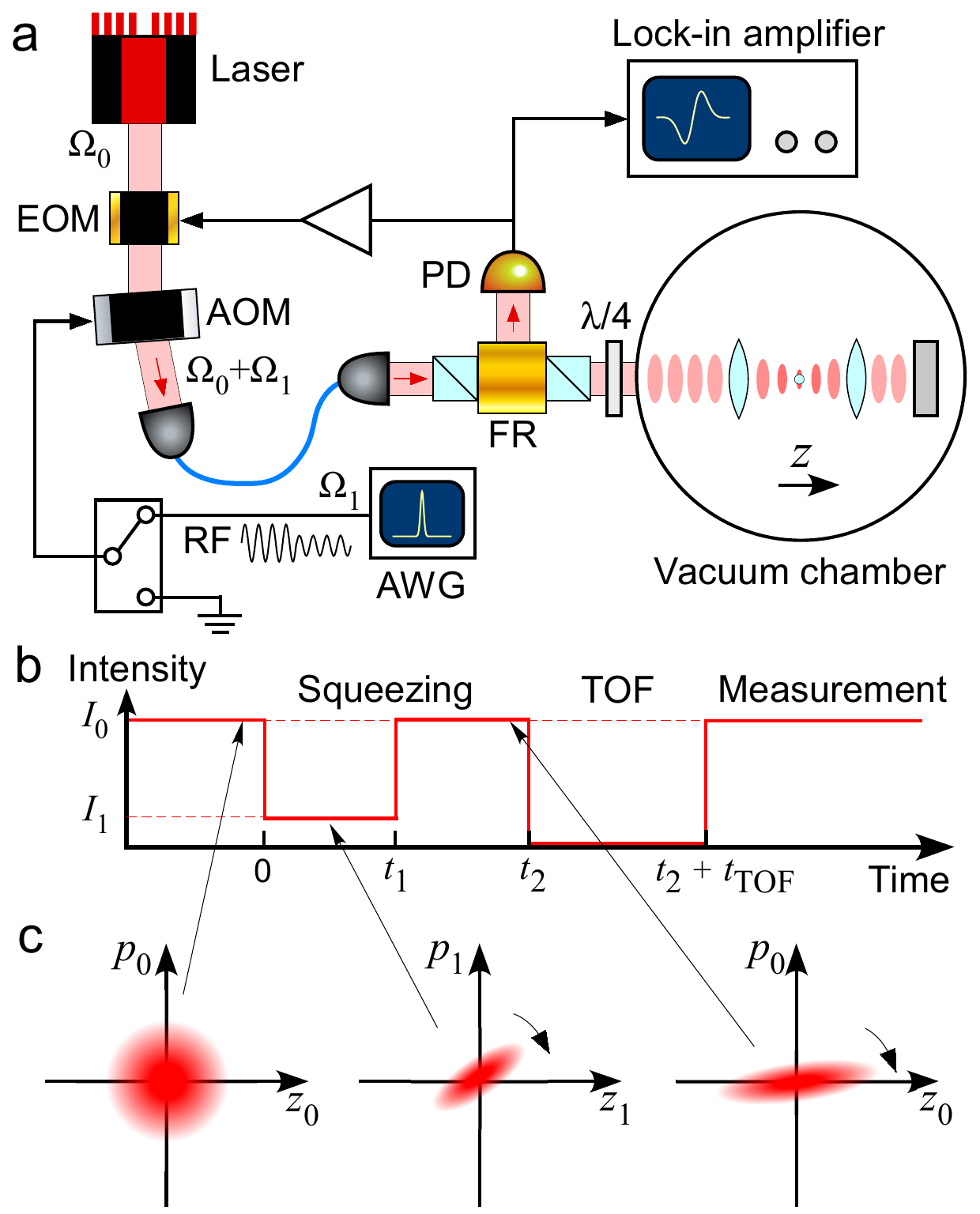} 
	\caption{\textbf{Experimental system.}
		({\bf a}) Brief schematic of the experimental setup. A single frequency laser at a wavelength of 1551.38~nm forms an optical lattice in the vacuum chamber. The light scattered by the nanoparticle is collected through a Faraday rotator (FR) and is incident on a photodetector (PD). The signal from the PD is used for feedback cooling via modulating the laser phase with an electro-optic modulator (EOM). The intensity of the laser is controlled by an acousto-optic modulator (AOM) driven by radio frequency (RF) from an arbitrary waveform generator (AWG). The light is turned off by switching off the RF with a switch. A beam for cooling motions perpendicular to the $z$ direction is incident from the right side. ({\bf b})  Time sequence for the squeezing protocol.  ({\bf c}) Variation of the uncertainty on the normalized phase space during the squeezing protocol. $z_i,p_i$ indicates the position normalized by $\sqrt{\hbar/2m\omega_i}$ and the momentum normalized by $\sqrt{\hbar m \omega_i/2}$, respetively, with $i=0,1$. }
	\label{fig:expset} % give each figure a logical label name
\end{figure}

Oscillators have played a crucial role in diverse area of modern quantum technologies, ranging from clocks~\cite{takamoto2005optical,bloom2014optical} and sensors~\cite{degen2017quantum,pezze2018quantum} to superconducting devices~\cite{clarke2008superconducting,devoret2013superconducting}, where the accurate and precise measurement of the oscillation has been a crucial issue. Extensive studies with microscopic particles, most of which behave as harmonic oscillators, have revealed that the fundamental limit of observing the oscillation is imposed by the quantum mechanics as $\Delta z \cdot \Delta p \geq \hbar/2$, where $\hbar$, $\Delta z$, and $\Delta p$ are the reduced Planck's constant, the uncertainties in the position and the momentum, respectively. This uncertainty principle dictated by Heisenberg suggests that, even though both of quadrature components cannot be observed with infinite precision at the same time, either of them can be measured more precisely than that of the quantum ground state. Due to the absence of its classical analog, such a process is called quantum squeezing and has been a target of intensive researches aiming at enhancing the sensitivity of observing the oscillators~\cite{ma2011quantum,toth2014quantum,braun2018quantum}. Quantum squeezing also constitutes an important building block for quantum information processing~\cite{braunstein2005quantum}.  Bringing the knowledge achieved with microscopic particles to larger, more classical objects is an important task to extend the understanding and applications of quantum physics~\cite{barzanjeh2022optomechanics}.

A levitated nanomechanical oscillator, which is a solid particle confined in an optical or an electrical harmonic potential, provide an ideal opportunity for testing quantum mechanics at meso- and macro-scopic scales~\cite{millen2020quantum,millen2020optomechanics,gonzalez2021levitodynamics}. Being isolated from other objects, levitated particles are expected to be ideal for sustaining fragile quantum states, while the wide tunability of the trapping potential enables various quantum protocols developed with microscopic particles, including a wavefunction expansion via a free fall. In recent years, significant technical developments have realized cooling the motions of levitated nanoparticles to their quantum ground state~\cite{delic2020cooling,magrini2021real,tebbenjohanns2021quantum,kamba2022optical,ranfagni2022two}. Yet, observing quantum mechanical behaviors of their motions still remain challenging. An unexplored issue, unique to levitated quantum systems, is whether the the minute quantum effect in one motional degree of freedom can be revealed without being affected by other degrees of freedom.

\begin{figure}
\centering
\includegraphics[width=0.9\columnwidth] {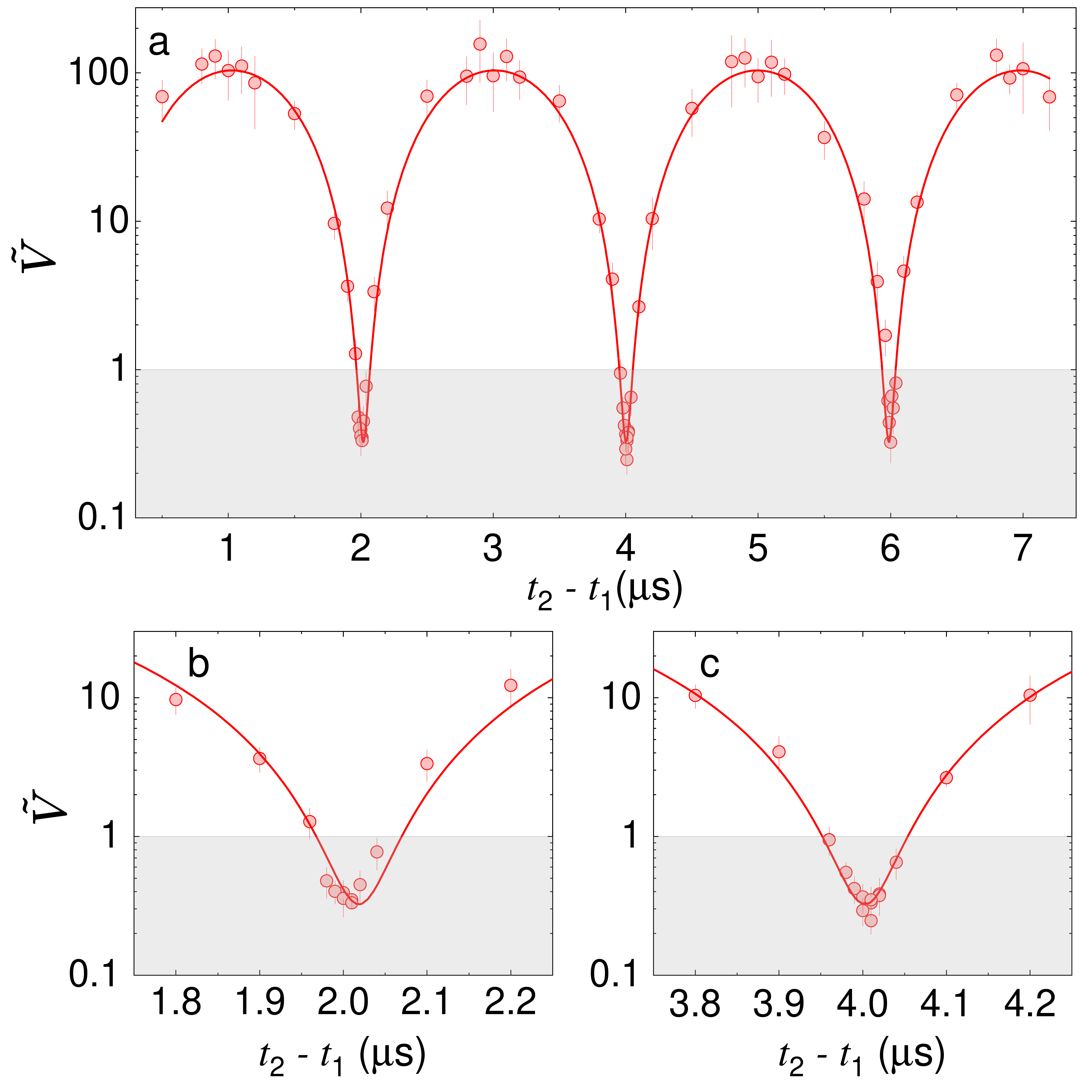}
\caption{ \textbf{Time evolution of the normalized velocity variance for $r=0.85$.}  ({\bf a}) Plot for the entire time range. When $t_2-t_1$ is varied, we observe a variation of the velocity variance for more than two orders of magnitude, implying the rotation of the elliptical uncertainty on the phase space with a period of about $4~\mu s$. The error bars indicate the error in determining the width with a fit to the measured velocity distribution. The shaded area indicates quantum squeezing. The observed minimum velocity variance obtained from a fit on the entire profile corresponds to the squeezing level of $-4.9(1)$~dB with respect to that of the ground state. ({\bf b,c}) Expanded views of {\bf a} around the minima of $\tilde{V}$. }
\label{fig:time}
\end{figure}

In the present study, we demonstrate quantum squeezing of the center-of-mass (CoM) motion of a levitated neutral fast-spinning nanoparticle by the rapid modulation of the laser intensity. We reveal a dramatic reduction of the velocity uncertainty directly via time-of-flight (TOF) free-expansion measurements. The observed velocity variance is narrower than that of the ground state by 4.9~dB, which is comparable to the largest mechanical quantum squeezing obtained thus far~\cite{lei2016quantum}. To quantitatively support our claim on the degree of squeezing, we propose and demonstrate a method to determine the displacement of the nanoparticle by utilizing an optical standing wave as a reference. This approach is in an excellent agreement with a conventional approach for calibrating the displacement and provides a quantitative confirmation of the measured level of squeezing.

Our experimental platform is a single charge-neutral nanoparticle levitated in the standing wave of a single-frequency laser, that is, an optical lattice~\cite{kamba2023nanoscale} (Fig.~\ref{fig:expset}a). Exploring a neutral nanoparticle has an advantage that its velocity can be measured via free expansion without the impact of residual electric field noises~\cite{kamba2023revealing}, similarly to the case with ultracold atoms~\cite{bloch2008many}. The optical lattice provides a high oscillation frequency and facilitates cooling to the ground state, while the strong feedback force required to reach the ground state is exerted by manipulating the lattice with laser phase modulation~\cite{kamba2022optical}.  
%The advantages of this system is the possibility to vary the trapping fruency to arbitraryly lower values, or even to turn off the trapping potential. The latter feature has been recently employed to reveal the profound correlation between the CoM motion near the ground state and the librational motions of a trapped nanoparticle~\cite{kamba2023revealing}. Despite the recent demonstrations of cooling its motion to the ground state~\cite{delic2020cooling,magrini2021real,tebbenjohanns2021quantum,kamba2022optical,ranfagni2022two},
The strong photon scattering via the nanoparticle is a key aspect enabling the continuous position measurement and feedback cooling, while it also imposes a relatively short motional coherence time with about 20 oscillations~\cite{delic2020cooling}. This short coherence time makes it challenging to realize squeezing via the parametric modulation of the potential, which has been successfully applied to achieve squeezing of up to -3~dB in various quantum systems~\cite{rugar1991mechanical,zalalutdinov2001optically,karabalin2010efficient,suh2010parametric}. Beating the -3~dB limit has been intensively explored both theoretically~\cite{clerk2008back,lu2015steady,agarwal2016strong} and experimentally~\cite{lei2016quantum}. 

%\subsection{Protocol for rapid squeezing}

To overcome the difficulty of short coherence time, we employ a scheme based on rapid variations of the oscillation frequency~\cite{xin2021rapid,wu2024squeezing,genoni2015quantum}.  In this scheme, a sudden decrease in the oscillation frequency is followed by the recovery of the frequency to the original value (Fig.~\ref{fig:expset}b).  Similar schemes have been useful for exploring thermal squeezed states~\cite{rashid2016experimental} and the delocalization of the nanoparticle's position~\cite{rossi2024quantum,muffato2024coherent}. We consider a phase space with the coordinates of the position normalized by $\sqrt{\hbar/2m\omega_0}$ and the momentum normalized by $\sqrt{\hbar m \omega_0/2}$, where $m$ and $\omega_0$ are the mass of the nanoparticle and the oscillation frequency, respectively. When the oscillation frequency is abruptly varied from $\omega_0$ to $\omega_1$, the uncertainty on the phase space is modified by a factor of $\exp(-r)$, where $r=\ln (\omega_0/\omega_1)/2=\ln (I_0/I_1)/4$ is a squeezing parameter~\cite{xin2021rapid}, with $I_0$ and $I_1$ being the initial intensity and the decreased intensity, respectively. The elliptical uncertainty rotates on the phase space at the frequency of oscillation (Fig.~\ref{fig:expset}c). In this scheme, both the intensity reduction at $t=0$ and the intensity increase at $t=t_1$ modify the uncertainty on the phase space. When the time duration for each process is carefully chosen to be $\omega_1 t_1 = \pi/2$ and $\omega_0 (t_2-t_1)=N \pi$, with $N$ being an integer, the momentum uncertainty at $t_2$ can be minimized to $e^{-2r}$ of that of the initial state.

In our experiment, we investigate the CoM motion of a neutral silica nanoparticle with a radius of $137(3)$~nm in an optical lattice~\cite{kamba2022optical,fnote10}. The position of the nanoparticle is continuously measured via monitoring photons scattered by the nanoparticle. The position signal is used for both feedback cooling of the CoM motions and the free-expansion measurement of the velocity distribution. We focus on the motion along the optical lattice ($z$ direction) which has the oscillation frequency of $\omega_0=2\pi \times 252$~kHz. The initial state before the squeezing protocol has the occupation number of $n_z=0.98(17)$. The impact of librational motions on the CoM motion is a great concern~\cite{kamba2023revealing}. To stabilize the angular motions via a gyroscopic effect, we let the nanoparticle spin around the $z$ axis at a frequency of about $5.6(2)$~GHz by introducing a circularly polarized light into the trapping laser~\cite{reimann2018ghz,ahn2018optically}. 
%The motions perpendicular to the optical lattice are cooled to the occupation number of around 30. In our previous study, we found that librational motions broaden the width of the velocity distributions. In this study, to avoid the impact of we  Due to a gyroscopic effect, the orientation of the nanoparticle is expected to be stable. 
During the entire measurements, the background pressure is kept at around $1.0\times 10^{-6}$~Pa such that the decoherence rate via background gas collisions $\Gamma_{\rm BG}=2\pi \times 0.10(1)$~kHz is negligibly smaller than the decoherence rate due to photon scattering $\Gamma_{\rm qba}= 2\pi \times 2.1(1)$~kHz.

After the initial state preparation, we turn off feedback cooling and apply the squeezing protocol by modulating the laser intensity (Fig.~\ref{fig:expset}b). In the rest of this article, we focus on the velocity instead of the momentum, because the velocity is directly obtained from the experiments. To measure the velocity $v_z$ via a free-expansion measurement, we turn off  the laser for the TOF of $t_{\rm TOF} = 51~\mu$s. The displacement during the TOF, $z$, is measured by recording the oscillation amplitude after recapturing the nanoparticle. We obtain the distribution of the velocity $v_z=z/t_{\rm TOF}$ by repeating the same procedure for about 300 times. Given that the velocity distribution follows the Maxwell-Boltzmann distribution, we extract the velocity width $\Delta v_z$ by fitting the distribution with a Gaussian function:
\begin{eqnarray}
f(v_z) = \exp \left[-(v_z-v_0)^2/2(\Delta v_z)^2 \right] 
\label{eq:dist}
\end{eqnarray}
The center of the velocity distribution $v_0$ reflects a displacement during the TOF due to various forces such as the projection of gravity along the optical lattice and a radiation pressure from a weak leakage light.

When we vary $t_2-t_1$, we observe a drastic time evolution of the velocity variance $(\Delta v_z)^2$. A plot of the time evolution of $\tilde{V}(t)=(\Delta v_z)^2/V_0$, where $V_{0} =  \hbar \omega_0/2m$ is the variance of the ground state, for the squeezing parameter of $r=0.85$ is shown in Figure~\ref{fig:time}. In this measurement, $t_1$ is chosen such that the velocity width is minimized at $t=t_1$. We observe a variation of $\tilde{V}(t)$ over two orders of magnitude with a period of $\pi/\omega_0$, implying that the elliptical uncertainty is rotating on the phase space. At $t_2-t_1=N \pi/\omega_0 $, we observe significant reductions in $\tilde{V}(t)$ (Fig.~\ref{fig:time}b,c). 

We achieve a comprehensive understanding on the observed time evolution profile by fitting the following expression to the measured data~\cite{rossi2024quantum}.
\begin{eqnarray}
\tilde{V}(t) = \tilde{V}_{1} \cos^2 \left[ \omega_0 (t_2-t_1) \right] +\tilde{V}_{2} \sin^2 \left[ \omega_0 (t_2-t_1) \right]
\label{eq:vpt}
\end{eqnarray}
where $\tilde{V}_{1}$ and $\tilde{V}_{2}$ characterize the uncertainty on the phase space. The increase in the variance due to photon scattering $\Gamma_{\rm qba}$ is negligible in the presented measurement duration. We find that our measurements are in good agreement with Eq.(\ref{eq:vpt}). From the fit, we determine the values of $\tilde{V}_{1}$ and $\tilde{V}_{2}$ to be $0.32(1)$ and $104(4)$, respectively. $\tilde{V}_{1}$ determined with this fit corresponds to -4.9~dB of quantum squeezing. Our results overcome the well-known limit of -3~dB obtainable via a standard parametric modulation technique~\cite{milburn1981production} and is comparable to the value achieved with quantum reservoir engineering in a membrane oscillator~\cite{lei2016quantum}.

The choice of $t_{\rm TOF}$ requires careful considerations on the uncertainties of both the position and the velocity. For a non-squeezed state, which is typical with cold atom experiments, the contribution of the position uncertainty before free expansion is smaller than that of the velocity uncertainty by $\omega_0 t_{\rm TOF}$. Thus, $t_{\rm TOF} \gg 1/\omega_0$ is sufficient to ensure that the measured distribution is dominantly due to the velocity uncertainty. By contrast, a squeezed state explored in this study has a broader position uncertainty and a narrower velocity uncertainty. The requirement for dominantly observing the velocity uncertainty is therefore more stringent as  $t_{\rm TOF} \gg e^{4r}/\omega_0$. In the present study, the contribution of the velocity uncertainty is expected to be 87~\% even at the largest squeezing parameter. 

\begin{figure}
\centering
\includegraphics[width=0.9\columnwidth] {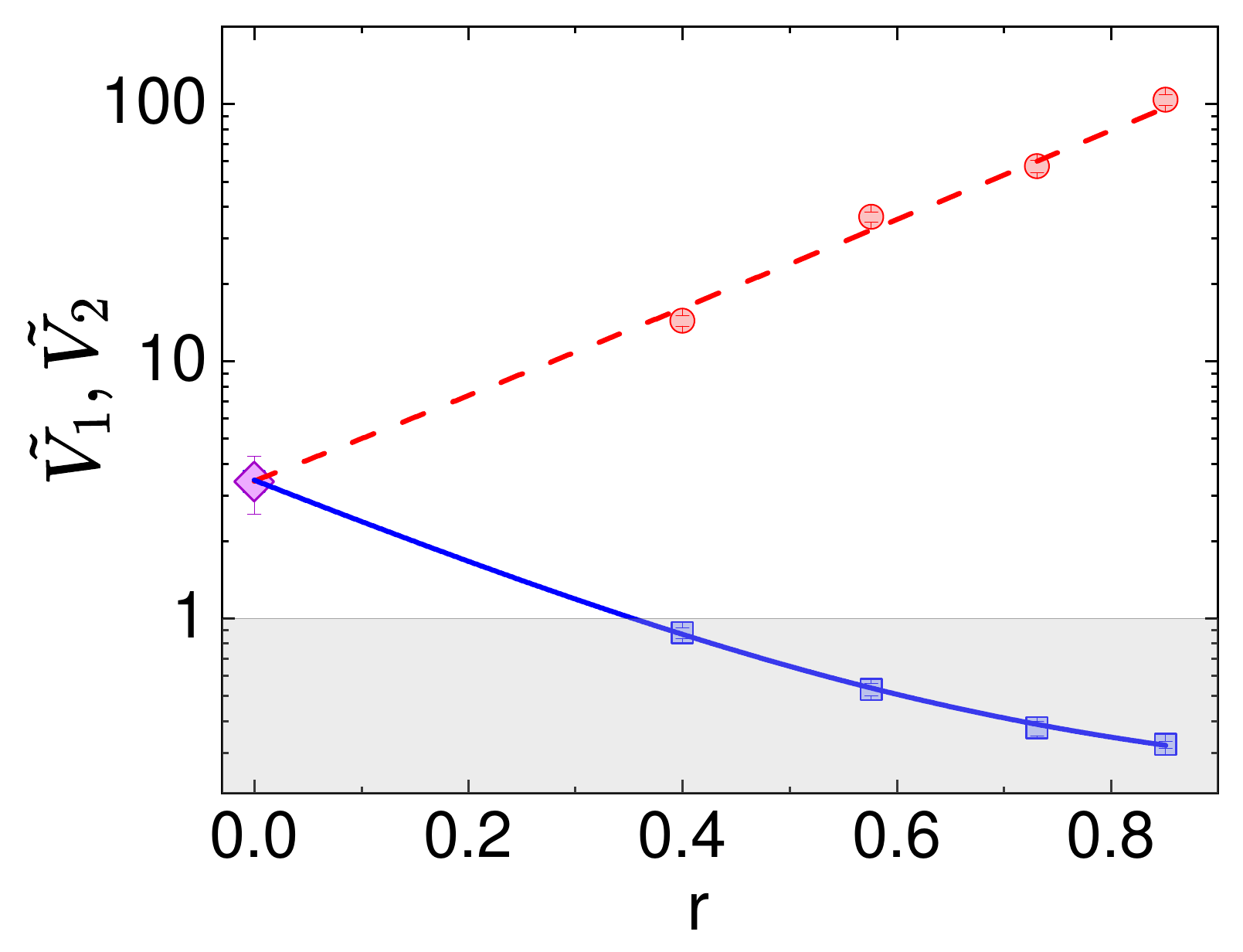}
\caption{ \textbf{Measured variance with respect to the squeezing parameter.} For each $r$, we plot $\tilde{V}_{1}$ (squares), $\tilde{V}_{2}$ (circles) extracted by fitting Eq.(\ref{eq:vpt}) to the measured time evolution. The variance of the initial state before the squeezing protocol is shown by a diamond at $r=0$. The error bars indicate the error in determining the parameters with a fit. The shaded area indicates quantum squeezing. From the fit on $\tilde{V}_{1}$ shown by a solid line, we find $\tilde{V_n}=0.21(1)$. For the fit on $\tilde{V}_{2}$ (a dashed line), we fix $\tilde{V_n}=0.21$. }
\label{fig:Vp}
\end{figure}

\begin{figure}
\centering
\includegraphics[width=0.9\columnwidth] {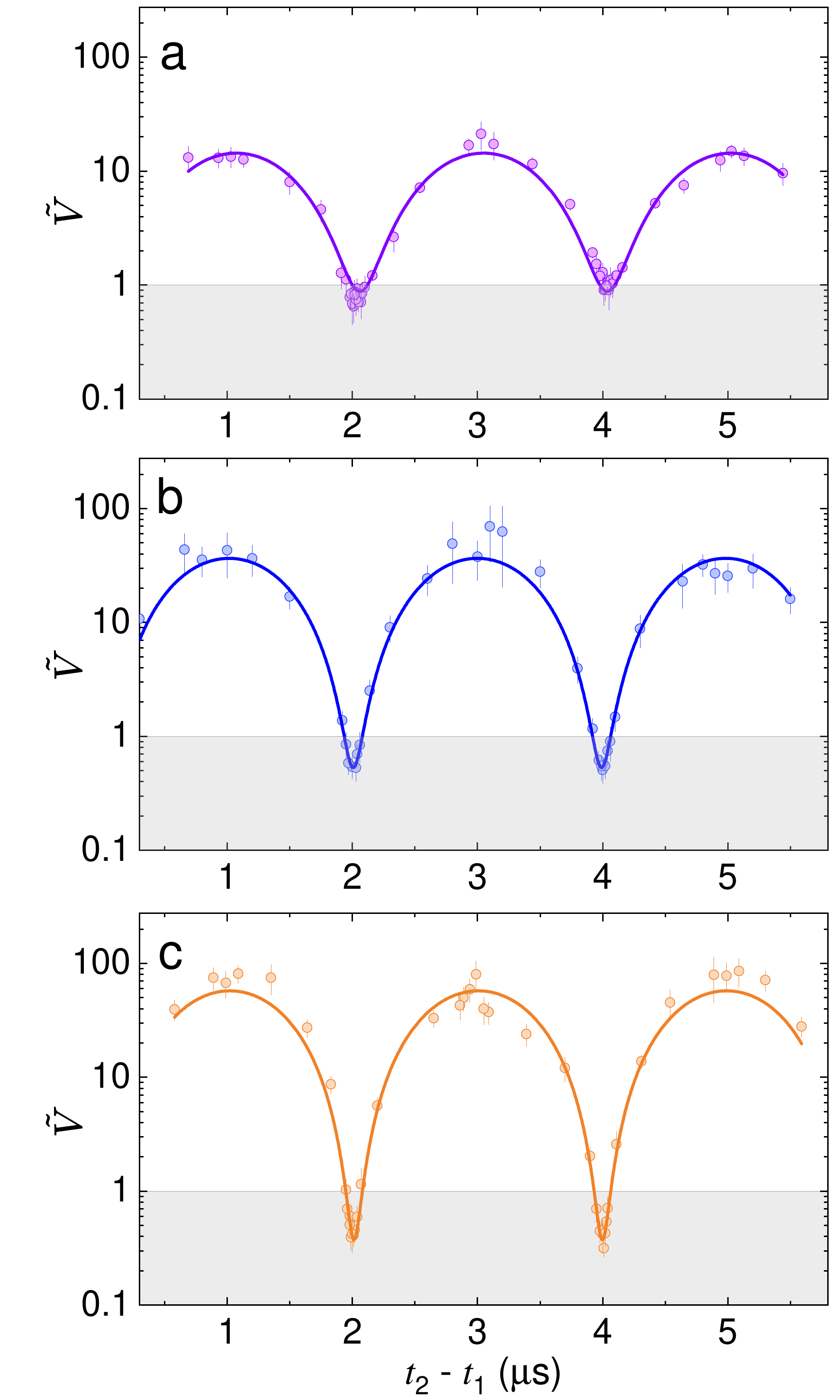}
\caption{ {\bf Time evolution of the velocity variance for various squeezing parameters.} ({\bf a}) $r =0.40$. ({\bf b}) $r =0.58$. ({\bf c}) $r = 0.73$. For each $r$, $t_1$ is optimized to reach a minimum velocity uncertainty at $t_1$. The shaded area indicates quantum squeezing. }
\label{fig:varVp}
\end{figure}

%\subsection{Comparison with numerical simulations}
The limit of the squeezing at larger values of $r$ in our protocol is an intriguing issue. By repeating the measurements described above for various values of $r$, we reveal the dependence on $r$. The entire time evolutions of $\tilde{V}(t)$ for various values of $r$ are provided in fig.~\ref{fig:varVp}.  In Fig.~\ref{fig:Vp}, we show $\tilde{V}_{1}$ and $\tilde{V}_{2}$ extracted from the time evolutions as a function of $r$. In addition to the values obtained with the squeezing protocol, the variance before the squeezing protocol, obtained by performing TOF measurements for the initial state, is also shown at $r=0$. We find that the the observed variance is in good agreement with the following expression~\cite{fnote10,xin2021rapid,rossi2024quantum}:
\begin{eqnarray}
\tilde{V}_{1} = \tilde{V}_n+\tilde{V}_{\rm ini} \exp(-4r) \\
\tilde{V}_{2} = \tilde{V}_n+\tilde{V}_{\rm ini} \exp(4r)
\label{eq:vp0}
\end{eqnarray} 
where $\tilde{V}_{\rm ini}$ denotes the normalized velocity variance of the initial state. $\tilde{V}_n$ originates from verious sources of uncertainties. From the fit on $\tilde{V}_{1}$, we obtain $\tilde{V}_n=0.21(1)$, which is consistent with estimatad contributions from technical noises (table~\ref{tab:systematic}, Supplementary Text). The agreement with a simple model for our squeezing protocol suggests that our protocol is working properly and the level of squeezing is ultimately limited to $\tilde{V}_n$. We find that the vibration of the system due to a scroll pump and a turbo-molecular pump broadens the width with $\tilde{V}_n \simeq 0.25$. Hence, the measurements presented in this work are carried out with these pumps being turned off. We point out that $\tilde{V}_n$ can in principle include an effect of quantum mechanical origin. When the measurements are performed simultaneously for both the position and the momentum, as realized with short $t_{\rm TOF}$ in our scheme, the quantum mechanical uncertainty of the detector can also come into play~\cite{arthurs1965bstj,yamamoto1986preparation,arthurs1988quantum}. There has been an argument that optically levitated nanoparticles in the weak measurement regime undergoes the simultaneous position and momentum measurements~\cite{rossi2024quantum}. 

\begin{figure}
\centering
\includegraphics[width=0.9\columnwidth] {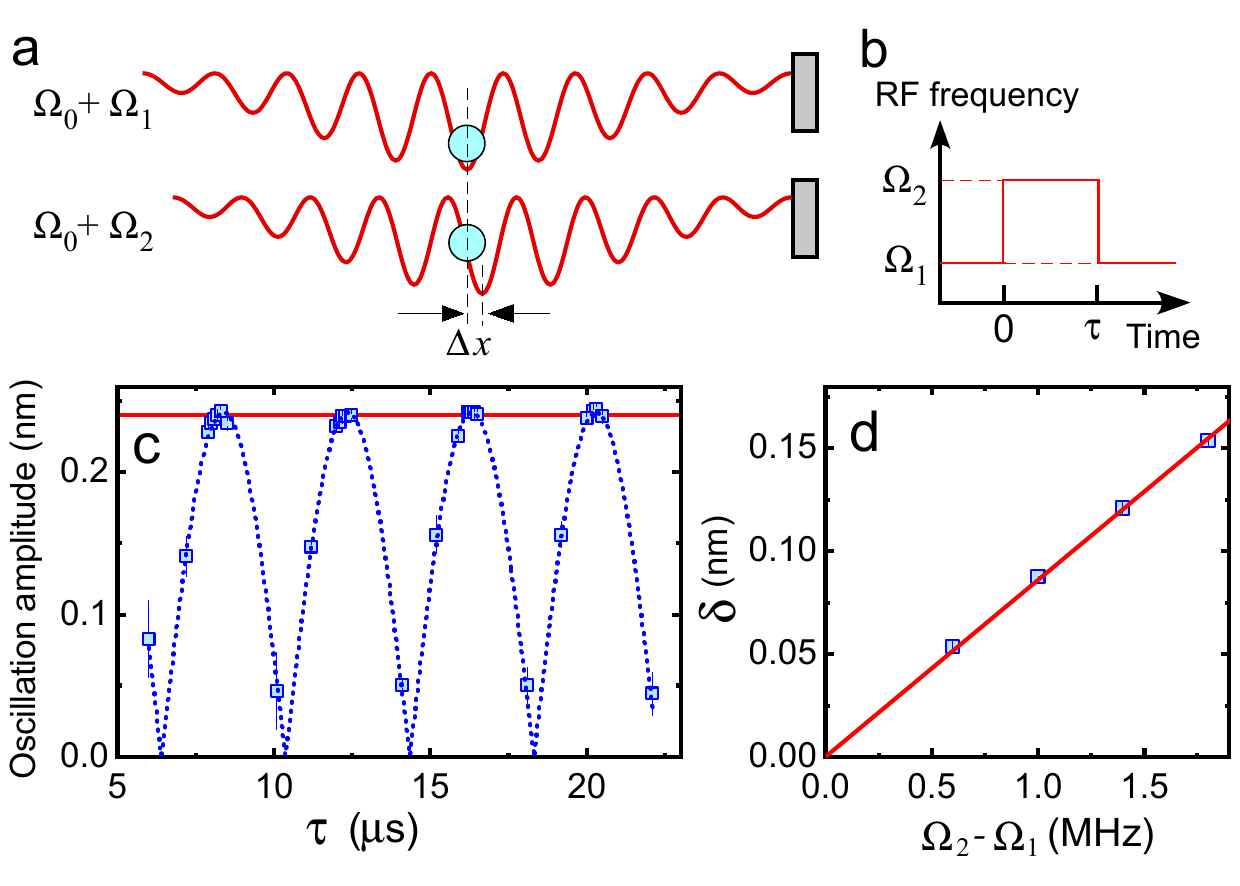}
\caption{ \textbf{ Calibration of the position signal via shifting the laser frequency. } ({\bf a}) Schematic showing the mechanism of inducing the oscillation of the trapped nanoparticle by shifting the laser frequency. ({\bf b}) Time sequence of the measurement. ({\bf c}) Oscillation amplitude with respect to $\tau$. The error bars indicate the standard deviation of the oscillation amplitude. The dashed lines is a fit of Eq.(\ref{eq:RFC}) to the measured amplitude, while the solid line is a calculated value with Eq.(\ref{eq:df}) and is not a fit.  ({\bf d}) Extracted amplitude with respect to the frequency difference. The error bars indicates the error in determining the amplitude with a fit to the oscillation signal. The solid line shows calculated values with Eq.(\ref{eq:df}) and is not a fit. }
\label{fig:RFC}
\end{figure}

Our claim of quantum squeezing relies on the basis of the correctly calibrated position signal because the level of squeezing is determined by taking the ratio of the measured velocity variance to $V_{0}$. All the plots presented in the present work relies on the following calibration procedure. We first perform TOF measurements of $\Delta v_z$ for various values of $n_z$ prepared by varying the feedback gain. By fitting the measured relation between $\Delta v_z$ and $n_z$ with the Maxwell-Boltzmann distribution, we find $\Delta v_z$ expected for $n_z=0$, which provides a calibration of the position signal (figure~\ref{fig:tof})~\cite{fnote10}. This approach is based on the value of $n_z$ obtained with the integration of power spectral density (PSD) under an assumption that the PSD without cooling thermalizes to room temperature~\cite{gieseler2012subkelvin,vovrosh2017parametric}. However, it has been argued that thermal fluctuations at room temperature can be a source of large errors~\cite{hebestreit2018calibration}. The flatness of the noise floor of the PSD is also an important prerequisite as the subtraction of the noise floor is required for an opt estimation of the area of the PSD. This requirement is also important with the quantum mechanical approach based on Raman sideband asymmetry~\cite{tebbenjohanns2020motional}.

To quantitatively confirm that our calibration is correct, we introduce an independent approach for calibrating the position signal that uses the nature of an optical lattice and does not rely on the integration of the PSD. Because the nanoparticle is trapped in an optical lattice, the effective distance between the nanoparticle and the retro-reflecting mirror $L$ is proportional to the wavelength of the laser $2\pi c/(\Omega_0+\Omega_1)$. This fact indicates that shifting the laser frequency to $\Omega_0+\Omega_2$ should induce an oscillation with an amplitude proportional to the frequency difference $\Omega_2 - \Omega_1$ (Fig.~\ref{fig:RFC}a). The induced displacement of the optical potential is given by 
\begin{eqnarray}
\delta = L ( \Omega_2 - \Omega_1) /(\Omega_0 + \Omega_2) 
\label{eq:df}
\end{eqnarray} 
In our measurements, after the initial state preparation, we turn off feedback cooling and  shift the optical potential for a short time $\tau$, during which we let the particle oscillate, and then we recover the original potential for position measurements (Fig.~\ref{fig:RFC}b). When we vary $\tau$, we observe a clear oscillation with an amplitude of $2\delta$ (Fig.~\ref{fig:RFC}b). The measured amplitudes calibrated with the temperature measurements are in excellent agreement with Eq.(\ref{eq:df}) within 2~\% (Fig.~\ref{fig:RFC}d), indicating that both calibration procedures are reliable. Due to its simplicity, the proposed approach for calibrating the position signal can also be a useful alternative means for future sensing applications.

In conclusion, we demonstrate quantum squeezing of the CoM motion of a levitated nanoparticle by manipulating the velocity uncertainty with rapid frequency variations. We reveal the reduction of the velocity uncertainty by directly observing the narrowing of the velocity distributions with TOF free-expansion measurements. The velocity variance is reduced to -4.9~dB of that of the quantum ground state. The observed level of squeezing is comparable to the mechanical squeezing realized with quantum reservoir engineering in a micromechanical oscillator~\cite{lei2016quantum}. We quantitatively support our finding by demonstrating a novel calibration procedure that relates the shift in the laser frequency and the amplitude of the nanoparticle's oscillation, which is in an excellent agreement with a conventional approach with the integration of the PSD. Our results represent an important milestone for the generation of non-classical states of the motion of levitated nanoparticles. We anticipate that introducing a nonlinear optical or electrical potential into the present setup will enable us to obtain various non-Gaussian states, which can be employed for further studies on non-classical states~\cite{neumeier2024fast}. 

From the practical point of view, levitated particles are expected as a promising system for realizing sensing devices beyond current technologies~\cite{monteiro2020force}. As demonstrated thus far with various quantum systems, quantum squeezing is a powerful technique to enhance the sensitivity~\cite{ma2011quantum,toth2014quantum,braun2018quantum,barzanjeh2022optomechanics}. The presented shceme is expected to be useful for future applications in acceleration sensing. In fact, our measurements are significantly influenced by the vibration of the vacuum pumps. Extending the limit of squeezing may allow us to reveal an otherwise undetectable effect, including the quantum mechanical uncertainty via the simultaneous position and momentum measurements~\cite{arthurs1965bstj,yamamoto1986preparation,arthurs1988quantum}. Furthermore, our system provides exciting possibilities of exploring thermodynamics under continuous quantum measurements and feedback~\cite{doherty2012quantum,yada2022quantum}.

The present work also has an important implication in the future experiments for revealing the quantum mechanical properties of mesoscopic and macroscopic objects~\cite{romero2011quantum,bassi2013models}. In most of the proposals to elucidate the quantum nature of the motions of levitated particles, the free-expansion of the wavefunction is required~\cite{romero2011optically,stickler2018probing}. The asymmetric geometry and the inhomogeneity of the material can yield couplings among mechanical degrees of freedom inside the trap, which can cause an unexpected, complicated expansion dynamics~\cite{kamba2023revealing}. In addition, during the free expansion, various sources of external fields can affect the motions~\cite{hebestreit2018sensing,rossi2024quantum}. Despite these concerns, the present work shows that a neutral spinning nanoparticle is a promising system for exploring its quantum mechanical nature via free-expansion measurements.

%Acknowledgment:\\
\begin{acknowledgments}
We thank M.\,Aspelmeyer, M.\,Ueda, B.\,Stickler, K.\,Funo, T.\,Sagawa, S.\,Sugiura, M.\,Kozuma, and T.\,Mukaiyama for fruitful discussions. We are grateful to S.\,Otabe for the experimental assistance. M.\,K. is supported by JST, the establishment of university fellowships towards the creation of science and technology innovation (Grant No. JPMJFS2112) and JSPS (Grant No. JP24KJ1058). N.\,H is supported by JST, SPRING (Grant No. JPMJSP2108). This work is supported by the Murata Science Foundation, the Mitsubishi Foundation, the Challenging Research Award, the 'Planting Seeds for Research' program, Yoshinori Ohsumi Fund for Fundamental Research, and STAR Grant funded by the Tokyo Tech Fund, Research Foundation for Opto-Science and Technology, JSPS KAKENHI (Grants No. JP16K13857, JP16H06016, JP19H01822, and JP22K18688), JST PRESTO (Grant No. JPMJPR1661), JST ERATO (Grant No.JPMJER2302), JST CREST (Grant No. JPMJCR23I1), and JST COI-NEXT (Grant No. JPMJPF2015).
\end{acknowledgments}

\section{Supplementary information}
\subsection{Optical setup for the initial state preparation}

A single-frequency laser at a wavelength of 1551.38~nm and with a power of 215~mW is focused with an objective lens (NA$=0.85$) and is approximately quarter of the incident power is retro-reflected to form a standing-wave optical trap (an optical lattice).  We load nanoparticles by blowing up silica powders placed near the trapping region with a pulsed laser at 532~nm at pressures of about 400~Pa. At around 350~Pa, we apply a positive high voltage to induce a corona discharge and provide a positive charge on the nanoparticle. Then we turn on optical feedback cooling for the translational motions and evacuate the chamber. We neutralize the nanoparticle via an ultraviolet light at around $2 \times 10^{-5}$~Pa. The motion along the optical lattice ($z$ direction) is cooled via optical cold damping realized with  the laser phase modulation~\cite{kamba2022optical}. The motions in the $x$ and $y$ directions are cooled via optical cold damping realized with the intensity modulation of an independent beam with a power of 0.4~mW.

%\begin{figure}[t]
%\includegraphics[width=0.95\columnwidth] {Fig_TOF_nrep02.eps}
%\caption{ Velocity width with respect to number of data points.  The widths of the velocity distributions are extracted from histograms with various data points. The error bars indicate statistical errors in fitting the distribution.  }
%\label{fig:nrep}
%\end{figure}

\subsection{Estimation of the mass and the temperature of nanoparticles}

We estimate the density and the radius of the trapped nanoparticle via the two independent measurements. First, we measure the heating rate at around 5~Pa, which is given by the background gas collisions~\cite{gieseler2012subkelvin,vovrosh2017parametric,iwasaki2019electric}. We measure the pressure with an accuracy of 0.5~\% via a capacitance gauge. Second, we measure the heating rate at around $1.0 \times 10^{-6}$~Pa, which is determined dominantly by photon recoil heating and is more sensitive to the radius than the heating rate at 5~Pa. By combining these results, we determine the radius and the density of the nanoparticle to be $137(3)$~nm and $2.26(3) \times 10^3$~kg/m$^3$, respectively. The mass of the nanoparticle is $m=2.4(2)  \times 10^{-17}$~kg.

The CoM temperatures are obtained by comparing the areas of the PSDs with and without cooling, as has been performed in previous studies~\cite{gieseler2012subkelvin,vovrosh2017parametric,iwasaki2019electric}. To avoid the influence of the increase in the internal temperature of nanoparticles at high vacuum due to laser absorption~\cite{hebestreit2018measuring}, we take the uncooled data at around 5~Pa. The PSD is averaged over 400 times to minimize thermal fluctuations. In addition, we take more than 10 averaged traces to derive the mean value of the area of the PSD as well as its standard deviation. We find that the typical thermal fluctuation of the area of the PSD is lower than 5~\% for both cooled and uncooled data. Thus, we estimate the systematic error in determining the temperatures of CoM motions to be about 7~\%. 

\subsection{Suppression of librational motions via spinning the particle}

Due to a gyroscopic effect, librational motions of a fast-spinning particle is expected to be significantly suppressed. We optically rotate the particle via the following procedure. First, at around 5~Pa, we rotate the quarter waveplate placed between the FR and the vacuum chamber by 30~$^\circ$. After that, we turn on feedback cooling and decrease the pressure. The particle starts to rotate at pressures between 1~Pa and 0.01~Pa. We then gradually recover the linear polarization. During the entire measurement, the quarter waveplate is kept rotated by 1.5$^\circ$ from the angle for linear polarization to sustain the particle rotation.

We are able to observe the signal of the particle rotation only between 0.01~Pa and $4\times 10^{-4}$~Pa. From the measured rotation rates at low vacuum, we estimate the rotation rate at our working pressure to be about $5.6(2)$~GHz under an assumption that rotation rate is inversely proportional to the pressure. 

In our previous work, we showed that feedback cooling of all the librational motions significantly reduces the velocity uncertainty~\cite{kamba2023revealing}. In comparison with feedback cooling on librational motions, where we need to adjust the conditions for cooling at each free-expansion cycle, the rotation of the nanoparticle allows us to acquire the data at two orders of magnitude higher rates and is more suitable for repeating many cycles.

\subsection{Control of time sequences}
We control the entire time sequences with a timing jitter of less than $\pm$10~ns with pulse generators (Quantum composer, 9214). Each cycle of the measurements is repeated at a rate of 3~Hz. The intensity modulation via an AOM takes about 250~ns, which is more than 10 times faster than the oscillation period of $4~\mu$s.

\subsection{Data acquisition and analysis for velocity distributions}
To measure the velocity of the nanoparticle after free expansion, we record the photodetector signal after passing through two fourth-order high-pass filters (cutoff frequencies of 105~kHz and 150~kHz) with a lock-in amplifier (MFLI, Zurich instruments). These filters are required to remove low frequency noises. The number of repetition is about 300 in the present study, which is confirmed to be sufficient for deriving the reliable velocity width.

We perform the following data processing for each single measurement in order to further suppress low frequency noises from the recorded data. The data is filtered through a 10th order finite-impulse-response filter centered at 253~kHz with a bandwidth of 20~kHz. The filtered trace is fitted with a sinusoidal function and its amplitude is extracted. By dividing the amplitude by the TOF of $51~\mu$s, we obtain the velocity of the nanoparticle for each measurement. 

In deriving histograms of the velocity distributions, the choice of the bin width can be arbitrary. We find that a too large value for the bin width results in a distribution with a broader width, while a too small value for the bin width gives a large error in determining the width via fitting with a Gaussian profile.  In the present work, we set the bin width to be $h=1.75\sigma/N^{1/3}$, which corresponds to the half of the Scott's choice, where $\sigma$ and $N$ denote the standard deviation of the data and the number of data, respecitvely. 

Assuming that the measured distribution follows the Maxwell-Boltzmann distribution, the velocity width $\Delta v_z$ obtained by fitting Eq.(\ref{eq:dist}) to the measured distribution is related to $n_z$ via the following expression:
\begin{eqnarray}
\Delta v_z = \sqrt{\hbar \omega_0 (n_z+1/2)/m}
\label{eq:vwidth}
\end{eqnarray}

\subsection{Calibration of the position signal with temperature measurements}
\begin{figure}
\centering
\includegraphics[width=0.9\columnwidth] {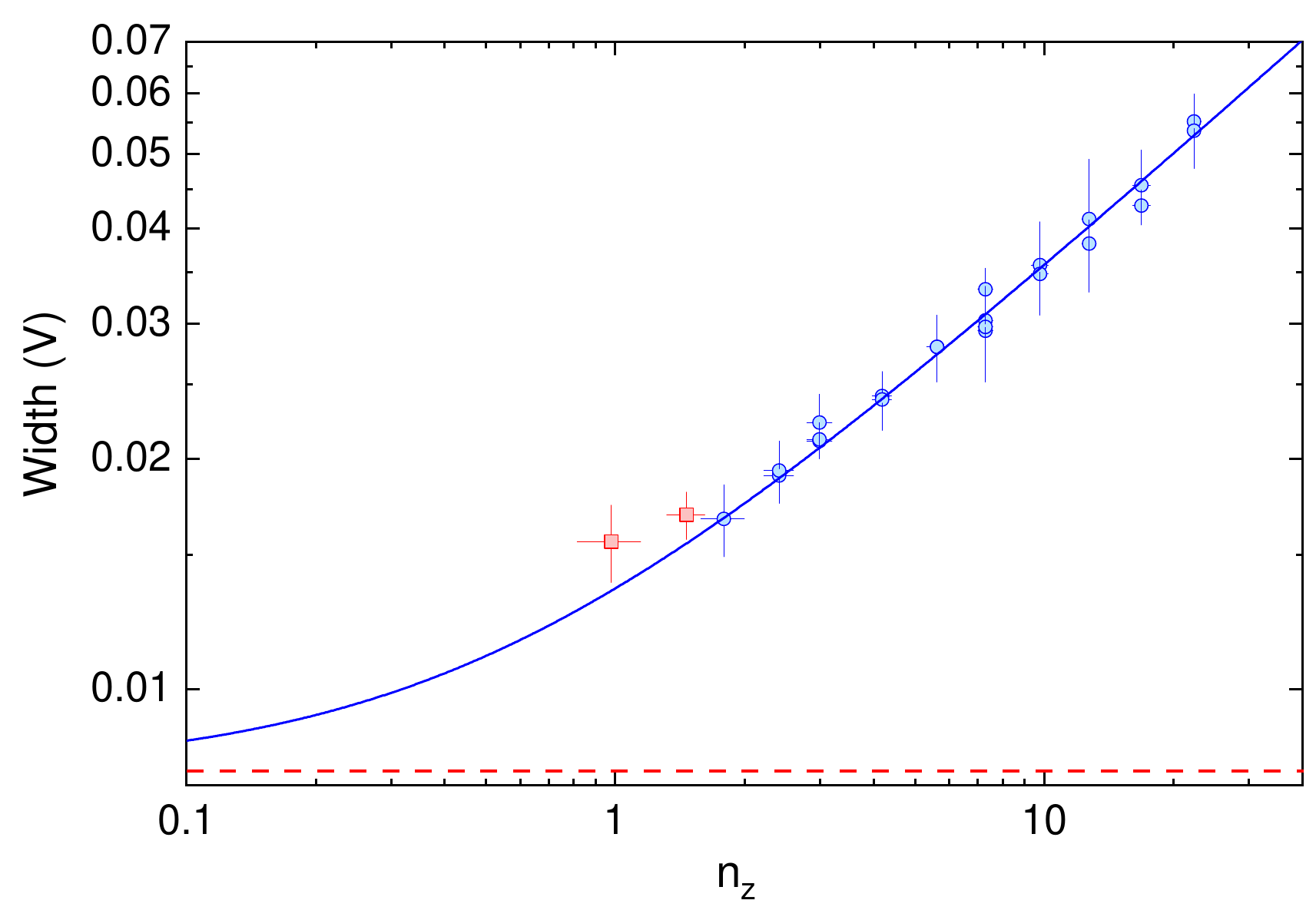}
\caption{ {\bf Velocity width with respect to the occupation number for the position calibration.} The velocity width in units of voltage is plotted with respect to the occupation number. The solid line is a fit with Eq.(\ref{eq:vwidth}) at $n_z>1.5$. From the fit on the measurements, we determine a relation between the occupation number and the width in voltage, which works as a position calibration. To exclude the effect of the slight broadening at low occupation numbers (square points), the fit is carried out in the range $n_z>1.5$. The width of the ground state is shown by a dashed line. }
\label{fig:tof}
\end{figure}

We employ the TOF measurements as a main means to find the relation between the voltage signal for the position measurements and the actual displacement. We first measure the temperature of the motion in the $z$ direction $T_z$ by taking the ratio of the area of the PSDs with and without feedback cooling. By varying the feedback gain, we are able to prepare temperatures between $18~\mu$K and $400~\mu$K. The lowest temperature corresponds to the occupation number of $n_z=0.98$, which is higher than our previous work because of the elevated imprecision in observing the position of a smaller nanoparticle.  We find that the observed widths of the velocity distributions $\Delta v_z$ are proportional to $\sqrt{n_z+1/2}$ at $n_z>1.5$ in accordance with the Maxwell-Boltzmann distribution (Fig.~\ref{fig:tof}). The deviation at low $n_z$ was reported in our previous study and can be attributed to a minute oscillation induced by turning off the feedback cooling. To avoid such an effect on our calibration, we use the data at $n_z>1.5$ for deriving the relation between $n_z$ and  $\Delta v_z$. 

\subsection{Calibration of the position signal by referencing an optical lattice}
The AOM for varying the laser intensity and turning off the laser is driven by an RF switch (ZASWA-2-50DRA+, Mini-circuits). For TOF free-expansion measurements, we turn off the RF with the switch during the TOF. By contrast, we feed the RF signal at a frequency $\Omega_2$ instead of turning off the RF for the calibration with the shift of the laser frequency. The variation of the laser intensity due to shifting the driving frequency is negligibly small for $\Omega_2-\Omega_1<2\pi \times 1$~MHz, while at $\Omega_2-\Omega_1\geq 2\pi \times 1.4$~MHz the intensity drops by about 3~\%. In our scheme, such a drop only results in the slightly varied oscillation frequency of $\omega_0'$ within a displaced potential and does affect the position calibration.  The device for recording the oscillation is the same as TOF measurements and squeezing measurements. We take about 120 traces for each time duration, from which the center of the distribution is extracted as a mean oscillation amplitude. 

The oscillation signal $z$ induced by the frequency shift (Fig.~\ref{fig:RFC}c) is fitted with the following equation to extract the amplitude of the oscillation $\delta$ :
\begin{eqnarray}
z=2\delta |\sin \dfrac{\omega_0' t}{2}|
\label{eq:RFC}
\end{eqnarray}
The length between the nanoparticle and the retro-reflecting mirror $L=16.6(3)$~mm is the sum of the working distance of the lens, the effective thickness of the lens right before the mirror, and the distance between the lens and a mirror measured by a caliper. The effective thickness of the lens is an optical path length taking into account the refractive index of the lens.  The effect of the variation of the Gouy phase on $\delta$ is numerically confirmed to be about 0.1~\% and is negligible.

\subsection{Derivation of the time dependence of the variance}
The squeezing operation induced by our protocol is represented by a squeezing operator $S(-2r)=\exp [-r \hat{a}_0^2 + r\hat{a}_0^{\dagger 2} ]$, where $\hat{a}_0, \hat{a}_0^\dagger$ are the annihilation and creation operators, respectively, while the squeezing parameter $r$ is defined as $r=\ln(\omega_0/\omega_1) /2$ with  $\omega_0, \omega_1$  being the initial and reduced frequency, respetively. This squeezing operator transforms the position and momentum operators $\hat{X}_{\rm ini}=\sqrt{\hbar/(2m\omega_0)} (\hat{a}_0+\hat{a}_0^\dagger )$ and $\hat{P}_{\rm ini}=-i \sqrt{m\hbar \omega_0/2} (\hat{a}_0-\hat{a}_0^\dagger )$ to $\hat{X}_{\rm sq}=e^{2r} \sqrt{\hbar/(2m \omega_0 )} (\hat{a}_0+\hat{a}_0^\dagger )$ and $\hat{P}_{\rm sq}=-i e^{-2r} \sqrt{m \hbar \omega_0/2} (\hat{a}_0-\hat{a}_0^\dagger )$, respectively.

%In order to characterize the uncertainty of the squeezed state at $t=t_1$, the velocity distribution of the nanoparticle is measured. At $t_2$, trapping laser is suddenly turned off to let the particle move freely according to its inertia, and after a TOF period of $t_{\rm TOF}$, it is recaptured by the optical trap to evaluate the deviation during the free flight time.
Since the particle oscillates in the original harmonic potential for $t_2-t_1$ before the release, the position and momentum operators just before the free flight are
\begin{eqnarray}
\hat{X}_{\rm TOF} (t_2 )=&\hat{X}_{\rm sq} \cos \left[ \omega_0 (t_2-t_1) \right] + \dfrac{\hat{P}_{\rm sq}}{m\omega_0} \sin \left[ \omega_0 (t_2-t_1) \right] \\
\hat{P}_{\rm TOF} (t_2 )=&-m\omega_0 \hat{X}_{\rm sq} \sin \left[ \omega_0 (t_2-t_1) \right] + \hat{P}_{\rm sq} \cos \left[ \omega_0 (t_2-t_1) \right]
\end{eqnarray}
On the assumption that the laser is turned off instantaneously, these operators directly give the initial state of the free flight. Then, the position and momentum operators just after the flight are given as
\begin{eqnarray}
\hat{X}_{\rm TOF} (t_2+t_{\rm TOF} )=&\hat{X}_{\rm TOF} (t_2 )+\dfrac{t_{\rm TOF}}{m} \hat{P}_{\rm TOF}(t_2) \\
\hat{P}_{\rm TOF}(t_2+t_{\rm TOF})=&\hat{P}_{\rm TOF}(t_2).
\end{eqnarray}
The kinetic energy after the recapture is given by
\begin{eqnarray}
E(t_2+t_{\rm TOF}) = \langle \dfrac{1}{2m}\hat{P}_{\rm TOF}^2 (t_2+t_{\rm TOF}) +\dfrac{1}{2}m\omega_0^2 \hat{X}_{\rm TOF}^2 (t_2+t_{\rm TOF} ) \rangle
\end{eqnarray}
Thus, the variance of the measured valocity normalized by $V_0$ is given by
\begin{eqnarray}
\tilde{V}(t)  =& \tilde{V}_{\rm ini} \Big\{ \dfrac{2\cosh(4r)}{(\omega_0  t_{\rm TOF})^2} - 2 \sin 2\left[ \omega_0 (t_2-t_1) \right] \dfrac{\sinh (4r)}{\omega_0 t_{\rm TOF} } \notag \\ 
 +& (e^{4r} \sin^2 \left[ \omega_0 (t_2-t_1) \right] +e^{-4r}  \cos^2 \left[ \omega_0 (t_2-t_1)  \right] \Big\}
\label{eq:var1}
\end{eqnarray}
The second term is vanishing at $t_2-t_1=N \pi $ with $N$ being an integer and can be ignored for deriving the level of squeezing with a fit. Furthermore, given that $e^{4r} \gg e^{-4r}$, Eq.(\ref{eq:var1}) is simplified as 
\begin{eqnarray}
\tilde{V}(t) =& \tilde{V}_1 \cos^2 \left[ \omega_0 (t_2-t_1) \right] +  \tilde{V}_2 \sin^2 \left[ \omega_0 (t_2-t_1) \right] \\ 
\tilde{V}_1 =& \tilde{V}_{\rm ini} \left( e^{-4r}+\dfrac{e^{4r}}{(\omega_0  t_{\rm TOF})^2} \right)\\ 
\label{eq:V1}
\tilde{V}_2 =& \tilde{V}_{\rm ini} \left( e^{4r}+\dfrac{e^{4r}}{(\omega_0  t_{\rm TOF})^2} \right) 
\end{eqnarray} 
The second term in $\tilde{V}_1,\tilde{V}_2$ indicates the contribution of the initial position uncertainty. In the present work, we include this term into $\tilde{V}_n$, which denotes various mechanisms for broadening the velocity distribution. Note that the achievable minimum variance decreases as $r$ gets larger. However, this decreasing trend is mitigated as $r$ increases, and the contribution of the initial position flucutation is comparable to the velocity fluctuation at $r= (\ln[1+(\omega_0 t_{\rm TOF})^2])/8$. 

\subsection{Estimation of the magnitude of various mechanisms that can contribute to $\tilde{V_n}$}
We estimate the contributions of various mechanisms that can broaden the measured velocity distribution. Due to a difficulty in calculations, the effect of the perpendicular motions is experimentally evaluated, while other mechanisms are considered by calculations based on previous studies and/or our measurements. The largest effect on $\tilde{V_n}$ was found to be the vibration of a turbo-molecular pump and a scroll pump that has been employed in our system for evacuation. The pumps can potentially shake the position of mirrors as well as a resonator for stabilizing the laser phase noise. Hence, in the present study, we turn off them and evacuate the system via a titanium-sublimation pump and an ion pump during the measurements.

\begin{table}
  \caption{Estimation of the contributions to $\tilde{V_n}$}
  \label{tab:systematic}
  \centering
  \begin{tabular}{lcr}
    \hline
    Source  & Experimental upperbound & Calculation \\
    \hline \hline
    (a)  & & $1.6 \times 10^{-2}$ \\
    (b)  & & $3.4 \times 10^{-5}$ \\
    (c)  &   & $1.0 \times 10^{-16}$ \\
    (d)   & & $8.6 \times 10^{-3}$ \\
    (e)  &  &  $4.1 \times 10^{-6}$ \\
    (f)  & $<6.4 \times 10^{-2}$ & \\
    (g)  &  &  $2.5 \times 10^{-2}$ \\
    (h)  &$<7.2 \times 10^{-2}$  &   \\
    \hline \hline
    Total & $<0.19$  \\
    \hline
  \end{tabular}
\end{table}

\subsubsection*{(a) Initial position uncertainty}
From Eq.(\ref{eq:V1}), the contribution of the initial position fluctuation is estimated to be $\tilde{V}_{\rm ini} e^{4r}/(\omega_0  t_{\rm TOF})^2 \simeq 1.6 \times 10^{-2}$ at $r=0.85$.

\subsubsection*{(b) Fluctuation of an optical lattice (near resonant)}
The residual laser phase noise can fluctuate the position of the particle through the fluctuation of an optical lattice. In our system, the laser phase noise is stabilized to $<1$~Hz/$\sqrt{\rm Hz}$. By multiplying $\lambda d/c$ to the phase noise density, where $\lambda = 1551.38$~nm and $d$ are the laser wavelength and the effective distance between the nanoparticle and the mirror, we find that the phase noise corresponds to the position fluctuation of $\Delta z \simeq 8.6 \times 10^{-17}$~m/$\sqrt{\rm Hz}$. The fluctuation of the velocity of the optical lattice near the resonant frequency $\omega_0$ is estimated to be $\Delta z \sqrt{2\pi/\omega_0}$, which gives a normalized variance of $3.4 \times 10^{-5}$. We ignore the effect of the displacement of the optical lattice during TOF as it is much smaller than this value.

\subsubsection*{(c) Slow drift of an optical lattice}
Apart from the near resonant phase noise considered above, the slow drift of the optical resonator used to stabilize the laser phase can fluctuate the optical lattice. The thermal expansion coefficient of our resonator, which is isolated in a vacuum chamber from the environment, is estimated to be about $10^{-8}$~/K, while we estimate the stability of the temperature within the typical duration of measuring a distribution, one minute, to be 10~mK. Hence, within the typical duration of measuring a distribution the drift of the resonator is estimated to be about 20~kHz. This frequency drifts indicates a velocity of $3 \times 10^{-14}$~m/s and corresponds to a normalized variance of $1 \times 10^{-16}$. The displacement of the optical lattice during TOF is an even smaller effect and is ignored here.

\subsubsection*{(d) Fulctuation of the angle of the optical table}
Through the measurements of the angle of the optical table, we find that it is tilted by about 2~$^\circ$ and a stability over a minute, a typical duration for taking a velocity distribution, is better than $\Delta \theta \simeq 0.1~^\circ $. Then the velocity fluctuation due to the fluctuation in the displacement during TOF is estimated to be $g t_{\rm TOF} \Delta \theta/2 \times \pi/180$, which corresponds to the normalized variance of $8.6 \times 10^{-3}$.

\subsubsection*{(e) Brownian motion of the reflecting mirror}
The Brownian motion of mirrors is a dominant mechanism that limits the stability of an optical resonator. Based on the value in Ref.~\cite{numata2004thermal}, we estimate the effect of brownian motion of the retro-reflecting mirror in our measurements. We assume that the position noise of the retro-reflecting mirror is $3 \times 10^{-17}$~m/$\sqrt{\rm Hz} $. Similarly to the fluctuation of the optical lattice near the resonant frequency, we consider only the fluctuation of the velocity of the optical lattice and estimate it to give a normalized variance of $4.1 \times 10^{-6}$.

\subsubsection*{(f) Effect of perpendicular motions}
We estimate the influence of motions perpendicular to the $z$ direction on $\tilde{V_n}$ by decreasing feedback gain for cooling them and increasing their temperatures. We find that the effect is less than $6.4 \times 10^{-2}$, which is limited by the error in deriving the width from the fit on the observed distribution. 

\subsubsection*{(g) Timing jitter}
Our experimental system is synchronized by using two pulse generators. The measured timing jitter of the time sequence is about $\pm 10$~ns. At the most squeezed condition in the present study, this timing jitter contributes to $\tilde{V_n}$ by $\tilde{V}_2 \sin^2 (\omega_0 \cdot 10~{\rm ns}) \simeq 2.5 \times 10^{-2}$.

\subsubsection*{(h) Residual vibration}
We measure the vibration of the optical table via an accelerometer (bandwidth $\approx 5$~kHz) with and without a scroll pump and a turbo-molecular pump. From this measurement, we find that the effect of the residual vibration with both pumps being turned off is less than $7.2 \times 10^{-2}$.
%The two main sections of the supplement can be split up using headings.

\bibliographystyle{apsrev}

%\bibliography{NPbib,ultracold}

\end{document}